\documentclass[twocolumn,prl,aps,showpacs]{revtex4-1}
\usepackage[T1]{fontenc}
\usepackage{xcolor}
\usepackage{amsthm}
\usepackage{amsmath}
\usepackage{amssymb}
\usepackage{graphicx}
\usepackage{subfig}
\usepackage{physics}

\begin{document}

\title{Optimized quantum state transfer through an XY spin chain}

\author{Yang Liu}

\affiliation{Beijing National Laboratory for Condensed Matter Physics,
  and Institute of Physics, Chinese Academy of Sciences, Beijing
  100190, China}

\author{D. L. Zhou}

\affiliation{Beijing National Laboratory for Condensed Matter Physics,
  and Institute of Physics, Chinese Academy of Sciences, Beijing
  100190, China}

\email{zhoudl72@iphy.ac.cn}

\begin{abstract}
  Quantum state transfer along a one-dimensional spin chain has become
  a fundamental ingredient for quantum communication between distant
  nodes in a quantum network. We study the average fidelity of quantum
  state transfer (QST) along a XY spin chain by adjusting the basis
  identification between the first spin and the last spin. In a proper
  choice of the basis identification, we find that the QST fidelity
  depends only on the average parity of the initial state linearly. We
  propose a simple scheme to adjusting the basis identification to
  optimize the average fidelity such that it depends linearly on the
  absolute value of the average parity. In the case that the absolute
  value of the average parity is $1$ we prove that the fidelity takes
  the maximum at any time over arbitrary initial state and basis
  identification.
\end{abstract}


\maketitle

\paragraph*{Introduction. ---}
A quantum wire that builds the communication channel between distant
nodes is a fundamental ingredient in a quantum network. The studies of
a spin chain as a quantum wire are pioneered by
Bose~\cite{PhysRevLett.91.207901}, where Bose showed that the high
fidelity of state transfer could be achieved through a long
unmodulated spin chain. Along an unmodulated spin chain, the perfect
state transfer is possible only when the length of the spin chain is
less than $4$. It is shown in
Refs.~\cite{PhysRevLett.92.187902,PhysRevA.71.032309,
  0295-5075-65-3-297,0953-8984-16-28-019} that the perfect state
transfer along an arbitrary long spin chain can be achieved by
modulating the coupling strengths. Along this direction, an exprement
in the framework of photonic lattices is reported to simulate the
mudulating coupling in Ref. \cite{Bellec:12}. The schemes that use the
spin chain without modulated coupling parameters in the limit of very
weak endpoint couplings are discussed in
Refs. \cite{PhysRevA.72.034303,PhysRevA.76.052328, PhysRevA.71.022301,
  1367-2630-12-2-025019, PhysRevA.78.022325, PhysRevLett.106.040505,
  PhysRevA.85.022312, PhysRevA.87.062309, PhysRevA.87.042313}, and the
optimization of one or two weak endpoint couplings is further
investigated in Refs. \cite{PhysRevA.85.052319,
  1367-2630-13-12-123006, PhysRevA.82.052321}. It is also shown that
it is possible to achieve perfect state transfer in a modulated spin
chain without initialization in
Refs.~\cite{PhysRevLett.101.230502,PhysRevA.79.054304}.

Recently Godsil \textit{et al} \cite{PhysRevLett.109.050502} proves a
beautiful result for QST along an XX spin chain in the single
excitation condition: XX spin chains can permit QST with a fidelity
arbitrarily closed to 1, if and only if the number of nodes is
$N=p-1,\,2p-1$, where $p$ is a prime, or $N=2^m-1$. However, numerical
results shows that the time to achieve a pretty good fidelity is very
long if $N$ is large.

All the above results motivate us to explore a more general problem:
For any given XY spin chain, what is the maximal QST fidelity it can
takes at a given time $t$? Notice that there exists two factors affect
the fidelity, one is the initial state of the spins except the sender,
the other is the basis identification between the first spin and the
last spin \cite{Bayat2011, 0295-5075-102-5-50003}. In other words, the central
task is how to optimally exploit an XY spin chain to QST at any given
time $t$.

The article is organized as follows. First, we will introduce the
model and explain the problem to be solved. Then we will show the
dynamics of the average fidelity of QST relates with only the dynamics
of observables for the output spin in the Heisenberg picture. Next we
will study the dynamics of these observables in the Heisenberg
picture, where a closed form is found for a general XY spin chain.
Form the closed forms, we obtain the general relation between the
fidelity and the parity of the initial state. Then the Laplace method
is used to solve the Heisenberg equation to obtain the dynamics of the
fidelity. Finally we propose a simple scheme to optimize the fidelity
for arbitrary initial states at any given time.

\paragraph*{Model and problem.---}
We consider a spin chain consisted of $N$ qubits, modeled by the XY
Hamiltonian:
\begin{equation*}
  H = \sum_{i=1}^{N-1}J_{i} X_{i} X_{i+1} + K_{i} Y_{i} Y_{i+1},
\end{equation*}
where $X_{i}$, $Y_{i}$ are the  Pauli matrices of spin $i$,
$J_{i}$ and $K_{i}$ are the coupling strengths. 

The process of quantum state transfer along the spin chain is as
follows. First, an unknown quantum state is prepared in the spin
labeled with $1$. Next we allow the unitary evolution controlled by the
Hamiltonian $H$ for a time period $t$. Then we check whether the
unknown state has been transferred to another spin labeled with $N$.

The quantity to characterize the QST process is the average fidelity
for an unknown state transferring from spin $1$ to spin $N$, which is
defined as
\begin{equation}
  \label{eq:1}
  F(t) = \int d\mu(\phi) \Tr(S \rho^{\phi}_{N} S^{\dagger} U(t)
  \rho^{\phi}_{1} \otimes \rho_{2\cdots N} U^{\dagger}(t)),\nonumber
\end{equation}
where $S$ is a unitary transformation on spin $N$,
$\rho^{\phi}=\op{\phi}$, $\rho_{2\cdots N}$ is the initial state of
the spins $2,3,\cdots,N$, and $U(t)$ is the unitary evolution of the
system. The introduction of $S$ means that we allow different
identifications of the basis vectors between spin $1$ and spin
$N$.

Obviously for a given Hamiltonian, the fidelity $F(t)$ depends on the
choices of $S$ and $\rho_{2\cdots N}$. The aim of this Letter is to
analyze how the choices of $S$ and $\rho_{2\cdots N}$ affect the
fidelity, and how to design the proper choices such that the fidelity
is optimized. Most importantly, what is the maximal fidelity among all
the choices? Is there a simple scheme to attain the maximal fidelity?

\paragraph*{Fidelity of QST in Heisenberg picture.---}
We start with the analysis of what need to be calculated to determine
the fidelity $F(t)$.

If we introduce a quantum channel
\begin{equation*}
  \mathcal{E}_{t}(\rho_{1}) = \Tr_{2\cdots{N}} ( U(t) \rho_{1} \otimes
  \rho_{2\cdots{N}} U^{\dagger}(t)),
\end{equation*}
then the fidelity $F(t)$ can be regarded as the fidelity between the
unitary transformation $S$ and the quantum operation
$\mathcal{E}_{t}$, which was simplified in
Refs.~\cite{BOS+2002,Nie2002}  as
\begin{equation}
  F(t) = \frac{1}{2} + \frac{1}{12} \sum_{\alpha\in\{X,Y,Z\}} \Tr(S
  \alpha_{N} S^{\dagger} \mathcal{E}_{t}(\alpha_{1})). \nonumber
\end{equation}
The meaning of the above equation is that the average over all
one-spin states can be reduced to the average over six states, namely
all the eigen-states of $X$, $Y$ and $Z$. Hence it makes the average
fidelity becomes observable in experiments.

Note that
$S\alpha_{N}S^{\dagger}=\sum_{\beta}R_{\alpha\beta}\beta_{N}$ with $R$
being a rotation matrix. Therefore we obtain
\begin{equation}
  \label{eq:3}
  F(t) = \frac{1}{2} + \frac{1}{12} \sum_{\alpha,\beta} R_{\alpha\beta}
  \Tr(\beta_{N}(t) \alpha_{1} \rho_{2\cdots N}),
\end{equation}
where $\beta_{N}(t)$ is the Pauli matrix $\beta_{N}$ in the Heisenberg
picture. Eq.~\eqref{eq:3} directly relates the fidelity $F(t)$ with
the Pauli matrices of spin $N$ in the Heisenberg picture. Thus to
obtain the fidelity $F(t)$ we only need to calculate the dynamics of
the Pauli matrices of spin $N$.

Because the time-dependent state in the Schrodinger picture contains
the dynamics of all the system's observables, the above result implies
that it possibly simplifies the study of the QST fidelity if we adopt
the Heisenberg picture other than the Schrodinger picture. We will
show it is indeed the case for the XY spin chain in the following.

\paragraph*{Fidelity and parity.---}
We start to  analyze the dynamics of $X_{N}(t)$, which satisfies the
Heisenberg equation
\begin{equation*}
  \label{eq:17}
  \dv{X_{N}}{t} = i \comm{H}{X_{N}}.
\end{equation*}

\begin{figure}[htbp]
  \centering
  \subfloat[$N=2M+1.$]
  { 
    \includegraphics{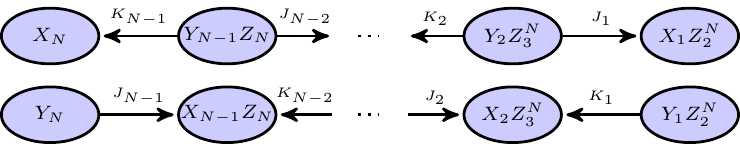}
  }
  
  \subfloat[$N=2M$.]
  {
    \includegraphics{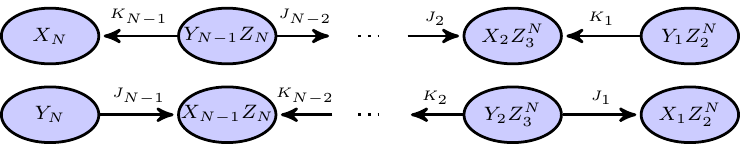}
  }
  \caption{\label{fig:Oriented-graph-describing} The sets of operators
    including $X_{N}$ or $Y_{N}$ that are closed under the action of
    $\comm{H}{\cdot}$. (a) The case when $N$ is odd. (b) The case when
    $N$ is even.}
\end{figure}

To solve the above equation, we first find the set of operators
including $X_{N}$ which is closed under the action
$\comm{H}{\cdot}$. We adopt the method given in
Ref.~\cite{PhysRevA.81.022319}, which is demonstrated in a graph shown
in Fig. \ref{fig:Oriented-graph-describing}. Every node in the graph
represents an operator. If we investigate the evolution of the
operator $\hat{O}$, for example $\hat{X}_N$ as we analyze in this
section, then put it in the first node. The node adjacent to the first
node is got by commuting first node with Hamiltonian $\hat{H}$. Other
nodes are got by the same way until we get all the elements of the
closed operators set. An outgoing (incoming) edge corresponds to a $+$
($-$) sign.

Hence we observed that the set of $N$ operators
\begin{equation}
  \label{eq:4}
  \Bqty{X_{N+2-2m}  Z_{N+3-2m}^{N}, \;
    Y_{N+1-2m}  Z_{N+2-2m}^{N}}
\end{equation}
with $m\in\{1,2,\cdots,[N]/2\}$ are closed under the action
$\comm{H}{\cdot}$, where $Z_{m}^{n}=\prod_{i=m}^{n}Z_{i}$. So
$X_{N}(t)$ can be expanded as
\begin{eqnarray}
  \label{eq:5}
  X_{N}(t) & = & \sum_{m} \left(a_{2m-1} X_{N+2-2m} 
    Z_{N+3-2m}^{N}  \right. \nonumber\\
  & & \left. +\, a_{2m} Y_{N+1-2m}  Z_{N+2-2m}^{N}\right).
\end{eqnarray}

Notice that the system is invariant under the transformation
\begin{equation}
  \label{eq:11}
  X_{n} \rightarrow Y_{n}, Y_{n} \rightarrow X_{n}, Z_{n} \rightarrow
  -Z_{n}, J_{i} \leftrightarrow K_{i} .
\end{equation}
Thus we have
\begin{eqnarray}
  Y_{N}(t) & = & \sum_{m} \left(b_{2m-1} Y_{N+2-2m} 
    Z_{N+3-2m}^{N}  \right. \nonumber\\
  & & \left. + b_{2m} X_{N+1-2m}  Z_{N+2-2m}^{N}\right),  
  \label{eq:12}
\end{eqnarray}
where
\begin{eqnarray}
  \label{eq:13}
  b_{2m-1}(t) & = & a_{2m-1}^{J_{i}\leftrightarrow K_{i}}(t), \\
  b_{2m}(t) & = &  - a_{2m}^{J_{i}\leftrightarrow K_{i}}(t).
\end{eqnarray}
Then $Z_{n}(t)$ can be obtained by $ Z_{N}(t)=-iX_{N}(t)Y_{N}(t)$.

From Eq.~\eqref{eq:3}, the parts of $X_{N}(t)$, $Y_{N}(t)$, and
$Z_{N}(t)$ that contribute to the fidelity $F(t)$ must contain
$X_{1}$, $Y_{1}$, or $Z_{1}$. So we can write $X_{N}$ and $Y_{N}$ into
two parts, one part that contributes to $F(t)$, and the other that
does not.  When $N$ is odd,
we get
\begin{equation*}
  \label{eq:15}
 X_{N}(t)  =   a_{N} X_{1} Z_{2}^{N} +\bar{X}_{N}, \quad
  Y_{N}(t)   =  b_{N} Y_{1} Z_{2}^{N} +\bar{Y}_{N}.
\end{equation*}
When $N$ is even, we have
\begin{equation*}
  \label{eq:16-1}
  X_{N}(t)  =   a_{N} Y_{1} Z_{2}^{N} +\bar{X}_{N}, \quad
  Y_{N}(t)   =  b_{N} X_{1} Z_{2}^{N} + \bar{X}_{N}.
\end{equation*}

According to the above results, it is reasonable to take $S=I$
($R_{\alpha\beta}=\delta_{\alpha\beta}$) when $N$ is odd, and
$S=\exp(i\frac{\pi}{4}Z_{N})$ ($R_{YX}=-R_{XY}=R_{ZZ}=1$ and all
other elements of $R$ are $0$) when $N$ is even.

Therefore we obtain the fidelity
\begin{equation}
  \label{eq:16}
  F(t) = \frac{1}{2} + \frac { \pqty{a_{N} +
      a_{N}^{J_{i}\leftrightarrow K_{i}}}  \ev{Z_{2}^{N}}  + a_{N}
    a_{N}^{J_{i}\leftrightarrow K_{i}} }  {6}.
\end{equation}

This implies that the fidelity of quantum state transfer is determined
by the average value of the parity for the initial state
$\rho_{2\cdots N}$ and the coefficient $a_{N}(t)$. For the initial
states $\rho_{2\cdots N}$ that have the same average value of parity,
the fidelity $F(t)$ will be the same.

\paragraph*{Structure of the solution of $a_{N}(t)$.---}
Now we come to the solution of $a_{N}(t)$, which is determined from
the Heisenberg equation for $X_{N}$:
\begin{equation}
  \label{eq:6}
  \dv{a}{t} = G a,
\end{equation}
where $a=(a_{1}(t),a_{2}(t),\cdots,a_{N}(t))^{T}$ with $T$ being the
transpose operation, $G$ is a tri-diagonal matrix
\begin{equation}
  \label{eq:10}
  G = \bmqty{0 & -2K_{N-1} & 0 & 0 & \cdots\\ 
    2K_{N-1} & 0 & 2J_{N-2}  & 0 & \cdots \\
    0 & -2J_{N-2} & 0 & -2K_{N-3} & \cdots\\
    0 & 0 & 2K_{N-3} & 0 & \cdots\\
    \vdots & \vdots & \vdots & \vdots & \vdots}.
\end{equation}
The initial condition is $a(0)=(1, 0, 0, \cdots, 0)^{T}$. Note that we
have the normalization condition $\sum_{i} a_{i}^{2}=1$ arising from
$\Tr(X_{N}^{2}(t))=2$. Since the matrix $G$ satisfies
$G^{\dagger}=-G$, Eq.~\eqref{eq:6} can be imagined as a Schrodinger
equation with the Hamiltonian being $iG$ in an $N$-dimensional Hilbert
space, which greatly reduces the computational complexity for our
problem.

When applying the Laplace transformation on Eq.~\eqref{eq:6}, we get
\begin{equation}
  \label{eq:2}
  A \bar{a}(p) = a(0),
\end{equation}
where $A=p-G$.

According to Cramer's rule in the basic matrix theory, we have
\begin{equation}
  \label{eq:8}
  \bar{a}_{N} = \frac{\det A_{N}^{(N)}}{\det A_{N}},
\end{equation}
where $A_{N}^{(N)}$ is the matrix $A$ whose $N$-th column vector
replaced by $a(0)$.

Notice that
\begin{equation}
  \label{eq:24}
  \det A_{N}^{(N)} =
  \begin{cases}
    (-1)^{M+1} \prod_{i=1}^{M}(2K_{2i-1}) \prod_{i=2}^{M} (2J_{2i-2})
    & \\ \hfill \text{if } N=2M,\\
    (-1)^{M} \prod_{i=1}^{M} (2K_{2i}) (2J_{2i-1}) & \\ \hfill  \text{if } N=2M+1.
  \end{cases}
\end{equation}
The aim is to find $\det A_{N}$, denoted as $F_{N}$. For the 
tri-diagonal matrix $A$, we have the iterative relation for its
determinant
\begin{subequations}
  \label{eq:9}
  \begin{eqnarray}
    F_{2m} & = & p F_{2m-1} + 4 K_{N-2m+1}^{2} F_{2m-2},\\
    F_{2m-1} & = & p F_{2m-2} + 4 J_{N-2m+2}^{2} F_{2m-3}.
  \end{eqnarray}  
\end{subequations}
The initial condition is $F_{-1}=0$ and
$F_{0}=1$. 

Then we can prove that
\begin{equation}
  \label{eq:14}
  F_{N} = 
  \begin{cases}
    \prod_{i=1}^{M}(p^{2}+q_{i}^{2}) & \text{if } N = 2M,\\
    p \prod_{i=1}^{M}(p^{2}+s_{i}^{2}) & \text{if } N = 2M+1.
  \end{cases}
\end{equation}
In general, we assume that $q_{i}\neq q_{j}$ and $s_{i}\neq s_{j}$ for
any $i\neq j$.  So the inverse Laplace transformation of $\bar{a}_{N}$
is
\begin{equation}
  \label{eq:18}
  \frac{a_{N}(t)}{\det A_{N}^{(N)}} = 
  \begin{cases}
    \sum_{i=1}^{M} \frac{\sin(q_{i} t)}{q_{i} \prod_{j\neq i}
      (q_{j}^{2}-q_{i}^{2})} & \text{for } N=2M,\\
    \sum_{i=0}^{M} \frac{\cos(s_{i}t)}{\prod_{j\neq i}
      (s_{j}^{2}-s_{i}^{2})} & \text{for } N=2M+1,
  \end{cases}
\end{equation}
where $s_{0}=0$.

Let us consider the special case where $J_{i}=\frac{1}{2}$, $K_{i}=\frac{K}{2}$.
Then we obatin $q_{k}=\sqrt{1+2K\cos\varphi+K^{2}}$, with $\varphi$
is the roots of the equation 
\[
\csc(\varphi)(K\sin(M\varphi+\varphi)-\sin(M\varphi))=0,
\]
 and $s_{0}=0$, $s_{k}=\sqrt{1+2K\cos\frac{k\pi}{M}+K^{2}}$, $k=1,\,2,\cdots M-1$.
When $K=1$, that is XX model, $q$ and $s$ are reduced to 
\[
q_{k}=s_{k}=2\cos(\frac{k\pi}{N+1}),k=1,2,\cdots,M.
\]
The detailed derivation of the above solution is given in the appendix.

\begin{figure}[htbp]
\includegraphics{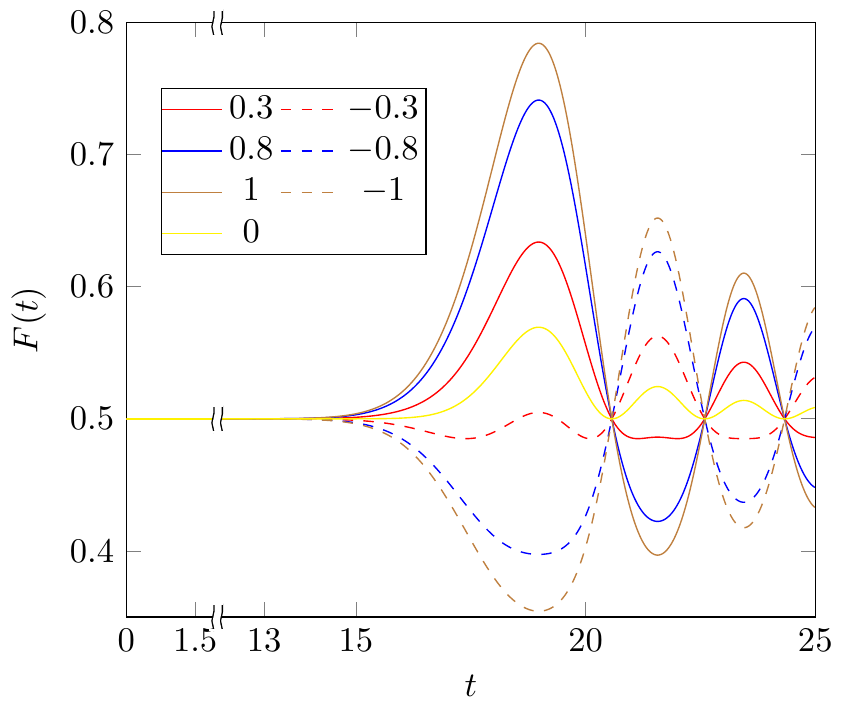}
\caption{The fidelity $F(t)$ for quantum state transfer in the XX
  model with $N=50$ for different parities $\ev{Z_{2}^{N}}$, that is
  $\ev{Z_{2}^{N}}=0$, $\pm 0.3$, $\pm 0.8$, $\pm 1$.}
\label{fig:fidepari}
\end{figure}
Now we demonstrate our results numerically. First, we show how the
parity affects the QST fidelity in Fig.~\ref{fig:fidepari} for the XX
model with $J=K=1$ and $N=50$. Before the time $t\simeq 14$,
$F(t)\simeq \frac{1}{2}$, which implies that the signal of the input
state propagates along the chain with the Lieb-Robinson velocity
\cite{Lieb1972,PhysRevLett.93.140402,Guo2012}. Obviously the fidelity
$F(t)$ oscillates with time for different parities, which reflects the
signal of the input state propagates in the spin chain. In general,
the fidelity for arbitrary initial state at any time is between the
one for the parity of $+1$ and that for the parity of
$-1$. Particularly, even the average parity is zero, i.e. for a
maximal mixed state, the fidelity might be larger than $1/2$.

\begin{figure}[htbp]
\includegraphics{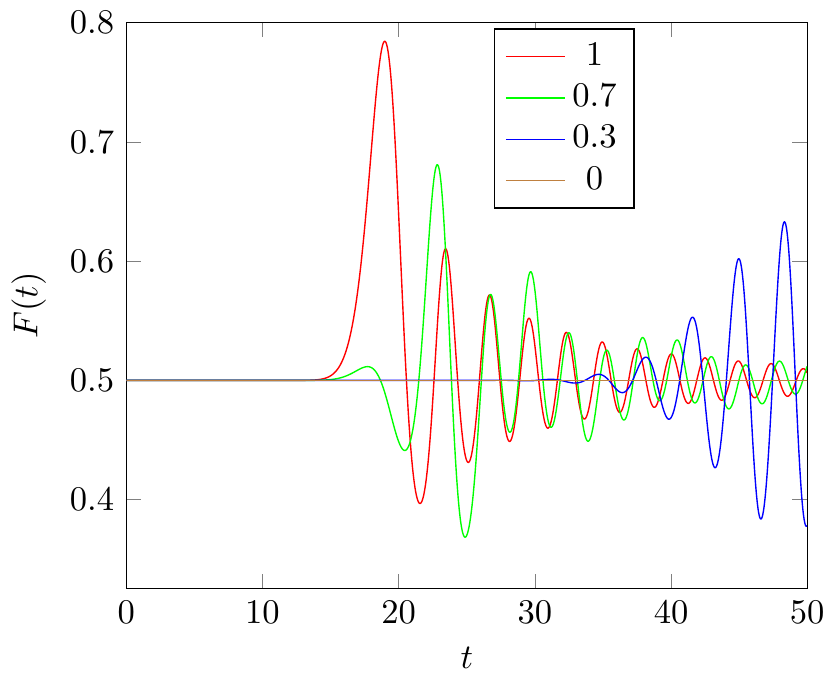}

\caption{The fidelity $F(t)$ for quantum state transfer in the
  different XY models with $N=50$, $J/K=1$, $0.7$, $0.3$, $0$.}
\label{fig:fideXY}
\end{figure}

Second, we demonstrate how the rate $J/K$ affects the fidelity $F(t)$
for a XY model in a spin chain with $N=50$ in
Fig.~\ref{fig:fideXY}. Note that the excitation number $\sum_{i}Z_{i}$
is not conserved in the case $J\neq K$. When $J=0$, the fidelity is
$1/2$ at any time, which implies that the signal of the input state
can never be transferred in this case. When $J/K$ is near $1$, the
behavior of the fidelity is similar to that of the XX model. However,
when $J/K$ is far from $1$, it shows a different oscillation behavior:
the delay start time and the increasing oscillation amplitude.

\paragraph*{Fidelity optimization by adjusting basis identification.---}

In Figs.~\ref{fig:fidepari} and \ref{fig:fideXY}, the fidelity
$F(t)<\frac{1}{2}$ for some $t$. As we know, the fidelity can reach
$\frac{1}{2}$ without any connection. Therefore, we can always make
the fidelity not less than $\frac{1}{2}$ by adjusting basis
identification. Here we emphasize that the basis identification needs
not any real operation, but only an agreement about the basis map
between the first spin and the last one.

Here we propose a simple scheme to adjust the unitary gate $S$. We
gives four unitary operations $i^{ab}Y_{N}^{a}X_{N}^{b}$ with
$a,b\in\{0,1\}$. Among the four choices, the optimized fidelity is
\begin{equation}
  \label{eq:20}
  F(t) = \frac{1}{2} + \frac { \pqty{\abs{a_{N}} +
      \abs{a_{N}^{J_{i}\leftrightarrow K_{i}}}}  \abs{\ev{Z_{2}^{N}}}
    +  \abs{a_{N} a_{N}^{J_{i}\leftrightarrow K_{i}}} } {6}.
\end{equation}

From Eq.~\eqref{eq:20}, if we increase the amplitude of the parity, we
can improve the fidelity $F(t)$. However, since the parity of spin $2$
to $N$ is a collective observable, we have no idea to increase its
absolute value to improve the fidelity by manipulating only the last
spin. In addition, Eq.~\eqref{eq:20} implies that the optimized
fidelity is not less than $\frac{1}{2}$ if we make a proper choice of
$a,b$.

Obviously, our optimized scheme can get the maximal fidelity when
$\abs{\ev{Z_{2}^{N}}}=1$, namely,
\begin{equation}
  \label{eq:21}
  F_{op}(t) =  \frac{1}{2} + \frac { \pqty{\abs{a_{N}} +
      \abs{a_{N}^{J_{i}\leftrightarrow K_{i}}}}  
    +  \abs{a_{N} a_{N}^{J_{i}\leftrightarrow K_{i}}} } {6}.
\end{equation}
Note that a similar result is  given in Ref. \cite{Bayat2011}
provided that $\abs{\ev{Z_{2}^{N}}}=1$. In fact, it is also the
maximal fidelity for arbitrary initial state and arbitrary unitary
transformation $S$, i.e., for any XY spin chain, we have
\begin{equation}
  \label{eq:22}
  F_{op}(t) = \max_{S,\rho_{2\cdots N}} F(t).
\end{equation}
The detail of the proof is given in the appendix.

\begin{figure}[htbp]
\includegraphics{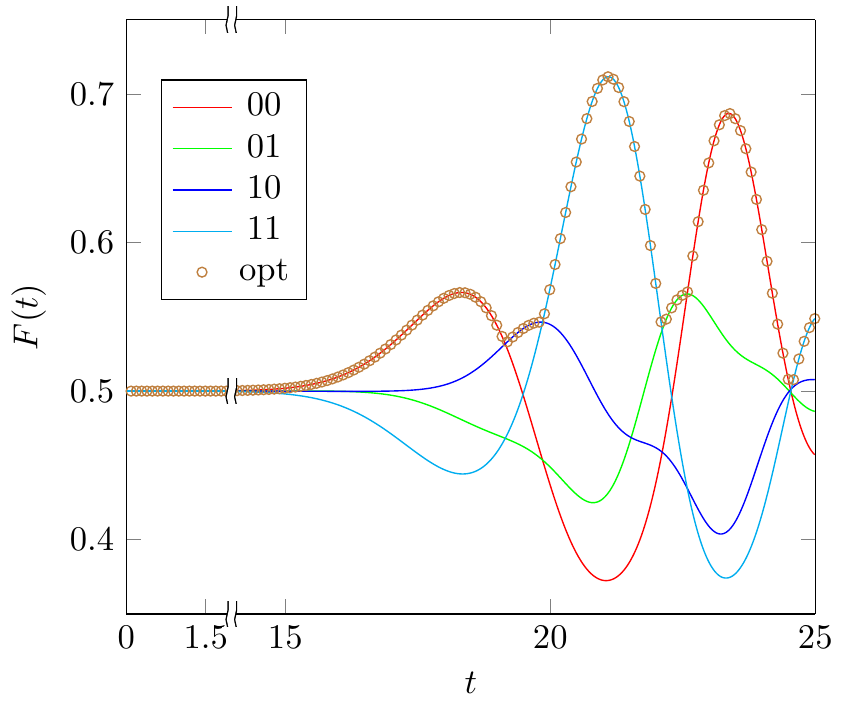}
\caption{The fidelity $F(t)$ for quantum state transfer in the XY
  models with $J/K=0.8$, $N=50$, $\ev{Z_{2}^{N}}=1$, but applied by
  different unitary gate $S$, that is $S=i^{ab}Y_{N}^{a}X_{N}^{b}$
  with $\{a,b\}=$ $\{0,0\}$, $\{0,1\}$, $\{1,0\}$, $\{1,1\}$.}
\label{fig:fideopti}
\end{figure}

We illustrate our optimized scheme of the QST fidelity in
Fig.~\ref{fig:fideopti} for the XY model with $N=50$, $J/K=0.8$, and
$\ev{Z_{2}^{N}}=1$. The optimized QST fidelity is a piecewise
function, where different unitary gates $S$ are taken in different
pieces.

\paragraph*{Discussion and conclusion.---}
Our result shows that the QST fidelity linearly depends on the parity
of the initial state. Then the maximal fidelity can be taken at the
parity $\ev{\sigma_{2}^{N}}=\pm{1}$. To achieve the perfect (or pretty
good) QST, it is sufficient to consider the case with the maximal
parity. In other words, if the perfect (or pretty good) QST can not
arrived at the initial state with the maximal parity, it also can not
arrived for arbitrary initial state. Notice that only the parity of
the initial state is relative, so when the initial state is in the
eigen-subspace of the parity, the QST fidelity will takes the maximum.

In the Schrodinger picture, when the initial state is complex or the
Hamiltonian does not conserve the excitation number, the QST will not
be equivalent to the propagation of the single
excitation. Eq.~\eqref{eq:6} implies that, however, the phenomena
similar to the single excitation propagation occurs in the Heisenberg
picture for any initial states. 

In summary, the analytical result on the QST fidelity along an XY spin
chain is given in the Heisenberg picture. It shows that the QST
fidelity only depends on the average value of the parity of the
initial states. We also propose a simple scheme to optimize the QST
fidelity by adjusting the basis identification with one of the
operations $I,X,Y,Z$ on the final spin, which ensures the QST fidelity
is not less than $\frac{1}{2}$ at any time for any initial states. We
prove that in terms of our optimizing scheme the maximal average
fidelity can be arrived when the amplitude of the average parity is
$1$. Therefore we gives a scheme on how to optimally use a XY spin
chain to transfer an unknown quantum state at any given time.

\begin{acknowledgments}
  We thank Leonardo Banchi and George Nikolopoulos for their helpful
  comments on the manuscript. This work is supported by NSF of China
  (Grant No. 11175247) and NKBRSF of China (Grant Nos.  2012CB922104
  and 2014CB921202).
\end{acknowledgments}

\bibliography{XX_information_flux}

\begin{thebibliography}{29}%
\makeatletter
\providecommand \@ifxundefined [1]{%
 \@ifx{#1\undefined}
}%
\providecommand \@ifnum [1]{%
 \ifnum #1\expandafter \@firstoftwo
 \else \expandafter \@secondoftwo
 \fi
}%
\providecommand \@ifx [1]{%
 \ifx #1\expandafter \@firstoftwo
 \else \expandafter \@secondoftwo
 \fi
}%
\providecommand \natexlab [1]{#1}%
\providecommand \enquote  [1]{``#1''}%
\providecommand \bibnamefont  [1]{#1}%
\providecommand \bibfnamefont [1]{#1}%
\providecommand \citenamefont [1]{#1}%
\providecommand \href@noop [0]{\@secondoftwo}%
\providecommand \href [0]{\begingroup \@sanitize@url \@href}%
\providecommand \@href[1]{\@@startlink{#1}\@@href}%
\providecommand \@@href[1]{\endgroup#1\@@endlink}%
\providecommand \@sanitize@url [0]{\catcode `\\12\catcode `\$12\catcode
  `\&12\catcode `\#12\catcode `\^12\catcode `\_12\catcode `\%12\relax}%
\providecommand \@@startlink[1]{}%
\providecommand \@@endlink[0]{}%
\providecommand \url  [0]{\begingroup\@sanitize@url \@url }%
\providecommand \@url [1]{\endgroup\@href {#1}{\urlprefix }}%
\providecommand \urlprefix  [0]{URL }%
\providecommand \Eprint [0]{\href }%
\providecommand \doibase [0]{http://dx.doi.org/}%
\providecommand \selectlanguage [0]{\@gobble}%
\providecommand \bibinfo  [0]{\@secondoftwo}%
\providecommand \bibfield  [0]{\@secondoftwo}%
\providecommand \translation [1]{[#1]}%
\providecommand \BibitemOpen [0]{}%
\providecommand \bibitemStop [0]{}%
\providecommand \bibitemNoStop [0]{.\EOS\space}%
\providecommand \EOS [0]{\spacefactor3000\relax}%
\providecommand \BibitemShut  [1]{\csname bibitem#1\endcsname}%
\let\auto@bib@innerbib\@empty
\bibitem [{\citenamefont {Bose}(2003)}]{PhysRevLett.91.207901}%
  \BibitemOpen
  \bibfield  {author} {\bibinfo {author} {\bibfnamefont {S.}~\bibnamefont
  {Bose}},\ }\href {\doibase 10.1103/PhysRevLett.91.207901} {\bibfield
  {journal} {\bibinfo  {journal} {Phys. Rev. Lett.}\ }\textbf {\bibinfo
  {volume} {91}},\ \bibinfo {pages} {207901} (\bibinfo {year}
  {2003})}\BibitemShut {NoStop}%
\bibitem [{\citenamefont {Christandl}\ \emph {et~al.}(2004)\citenamefont
  {Christandl}, \citenamefont {Datta}, \citenamefont {Ekert},\ and\
  \citenamefont {Landahl}}]{PhysRevLett.92.187902}%
  \BibitemOpen
  \bibfield  {author} {\bibinfo {author} {\bibfnamefont {M.}~\bibnamefont
  {Christandl}}, \bibinfo {author} {\bibfnamefont {N.}~\bibnamefont {Datta}},
  \bibinfo {author} {\bibfnamefont {A.}~\bibnamefont {Ekert}}, \ and\ \bibinfo
  {author} {\bibfnamefont {A.~J.}\ \bibnamefont {Landahl}},\ }\href {\doibase
  10.1103/PhysRevLett.92.187902} {\bibfield  {journal} {\bibinfo  {journal}
  {Phys. Rev. Lett.}\ }\textbf {\bibinfo {volume} {92}},\ \bibinfo {pages}
  {187902} (\bibinfo {year} {2004})}\BibitemShut {NoStop}%
\bibitem [{\citenamefont {Shi}\ \emph {et~al.}(2005)\citenamefont {Shi},
  \citenamefont {Li}, \citenamefont {Song},\ and\ \citenamefont
  {Sun}}]{PhysRevA.71.032309}%
  \BibitemOpen
  \bibfield  {author} {\bibinfo {author} {\bibfnamefont {T.}~\bibnamefont
  {Shi}}, \bibinfo {author} {\bibfnamefont {Y.}~\bibnamefont {Li}}, \bibinfo
  {author} {\bibfnamefont {Z.}~\bibnamefont {Song}}, \ and\ \bibinfo {author}
  {\bibfnamefont {C.-P.}\ \bibnamefont {Sun}},\ }\href {\doibase
  10.1103/PhysRevA.71.032309} {\bibfield  {journal} {\bibinfo  {journal} {Phys.
  Rev. A}\ }\textbf {\bibinfo {volume} {71}},\ \bibinfo {pages} {032309}
  (\bibinfo {year} {2005})}\BibitemShut {NoStop}%
\bibitem [{\citenamefont {Nikolopoulos}\ \emph
  {et~al.}(2004{\natexlab{a}})\citenamefont {Nikolopoulos}, \citenamefont
  {Petrosyan},\ and\ \citenamefont {Lambropoulos}}]{0295-5075-65-3-297}%
  \BibitemOpen
  \bibfield  {author} {\bibinfo {author} {\bibfnamefont {G.~M.}\ \bibnamefont
  {Nikolopoulos}}, \bibinfo {author} {\bibfnamefont {D.}~\bibnamefont
  {Petrosyan}}, \ and\ \bibinfo {author} {\bibfnamefont {P.}~\bibnamefont
  {Lambropoulos}},\ }\href {http://stacks.iop.org/0295-5075/65/i=3/a=297}
  {\bibfield  {journal} {\bibinfo  {journal} {EPL (Europhysics Letters)}\
  }\textbf {\bibinfo {volume} {65}},\ \bibinfo {pages} {297} (\bibinfo {year}
  {2004}{\natexlab{a}})}\BibitemShut {NoStop}%
\bibitem [{\citenamefont {Nikolopoulos}\ \emph
  {et~al.}(2004{\natexlab{b}})\citenamefont {Nikolopoulos}, \citenamefont
  {Petrosyan},\ and\ \citenamefont {Lambropoulos}}]{0953-8984-16-28-019}%
  \BibitemOpen
  \bibfield  {author} {\bibinfo {author} {\bibfnamefont {G.~M.}\ \bibnamefont
  {Nikolopoulos}}, \bibinfo {author} {\bibfnamefont {D.}~\bibnamefont
  {Petrosyan}}, \ and\ \bibinfo {author} {\bibfnamefont {P.}~\bibnamefont
  {Lambropoulos}},\ }\href {http://stacks.iop.org/0953-8984/16/i=28/a=019}
  {\bibfield  {journal} {\bibinfo  {journal} {Journal of Physics: Condensed
  Matter}\ }\textbf {\bibinfo {volume} {16}},\ \bibinfo {pages} {4991}
  (\bibinfo {year} {2004}{\natexlab{b}})}\BibitemShut {NoStop}%
\bibitem [{\citenamefont {Bellec}\ \emph {et~al.}(2012)\citenamefont {Bellec},
  \citenamefont {Nikolopoulos},\ and\ \citenamefont {Tzortzakis}}]{Bellec:12}%
  \BibitemOpen
  \bibfield  {author} {\bibinfo {author} {\bibfnamefont {M.}~\bibnamefont
  {Bellec}}, \bibinfo {author} {\bibfnamefont {G.~M.}\ \bibnamefont
  {Nikolopoulos}}, \ and\ \bibinfo {author} {\bibfnamefont {S.}~\bibnamefont
  {Tzortzakis}},\ }\href {\doibase 10.1364/OL.37.004504} {\bibfield  {journal}
  {\bibinfo  {journal} {Opt. Lett.}\ }\textbf {\bibinfo {volume} {37}},\
  \bibinfo {pages} {4504} (\bibinfo {year} {2012})}\BibitemShut {NoStop}%
\bibitem [{\citenamefont {W\'ojcik}\ \emph {et~al.}(2005)\citenamefont
  {W\'ojcik}, \citenamefont {\L{}uczak}, \citenamefont
  {Kurzy\ifmmode~\acute{n}\else \'{n}\fi{}ski}, \citenamefont {Grudka},
  \citenamefont {Gdala},\ and\ \citenamefont {Bednarska}}]{PhysRevA.72.034303}%
  \BibitemOpen
  \bibfield  {author} {\bibinfo {author} {\bibfnamefont {A.}~\bibnamefont
  {W\'ojcik}}, \bibinfo {author} {\bibfnamefont {T.}~\bibnamefont {\L{}uczak}},
  \bibinfo {author} {\bibfnamefont {P.}~\bibnamefont
  {Kurzy\ifmmode~\acute{n}\else \'{n}\fi{}ski}}, \bibinfo {author}
  {\bibfnamefont {A.}~\bibnamefont {Grudka}}, \bibinfo {author} {\bibfnamefont
  {T.}~\bibnamefont {Gdala}}, \ and\ \bibinfo {author} {\bibfnamefont
  {M.}~\bibnamefont {Bednarska}},\ }\href {\doibase 10.1103/PhysRevA.72.034303}
  {\bibfield  {journal} {\bibinfo  {journal} {Phys. Rev. A}\ }\textbf {\bibinfo
  {volume} {72}},\ \bibinfo {pages} {034303} (\bibinfo {year}
  {2005})}\BibitemShut {NoStop}%
\bibitem [{\citenamefont {Campos~Venuti}\ \emph {et~al.}(2007)\citenamefont
  {Campos~Venuti}, \citenamefont {Giampaolo}, \citenamefont {Illuminati},\ and\
  \citenamefont {Zanardi}}]{PhysRevA.76.052328}%
  \BibitemOpen
  \bibfield  {author} {\bibinfo {author} {\bibfnamefont {L.}~\bibnamefont
  {Campos~Venuti}}, \bibinfo {author} {\bibfnamefont {S.~M.}\ \bibnamefont
  {Giampaolo}}, \bibinfo {author} {\bibfnamefont {F.}~\bibnamefont
  {Illuminati}}, \ and\ \bibinfo {author} {\bibfnamefont {P.}~\bibnamefont
  {Zanardi}},\ }\href {\doibase 10.1103/PhysRevA.76.052328} {\bibfield
  {journal} {\bibinfo  {journal} {Phys. Rev. A}\ }\textbf {\bibinfo {volume}
  {76}},\ \bibinfo {pages} {052328} (\bibinfo {year} {2007})}\BibitemShut
  {NoStop}%
\bibitem [{\citenamefont {Li}\ \emph {et~al.}(2005)\citenamefont {Li},
  \citenamefont {Shi}, \citenamefont {Chen}, \citenamefont {Song},\ and\
  \citenamefont {Sun}}]{PhysRevA.71.022301}%
  \BibitemOpen
  \bibfield  {author} {\bibinfo {author} {\bibfnamefont {Y.}~\bibnamefont
  {Li}}, \bibinfo {author} {\bibfnamefont {T.}~\bibnamefont {Shi}}, \bibinfo
  {author} {\bibfnamefont {B.}~\bibnamefont {Chen}}, \bibinfo {author}
  {\bibfnamefont {Z.}~\bibnamefont {Song}}, \ and\ \bibinfo {author}
  {\bibfnamefont {C.-P.}\ \bibnamefont {Sun}},\ }\href {\doibase
  10.1103/PhysRevA.71.022301} {\bibfield  {journal} {\bibinfo  {journal} {Phys.
  Rev. A}\ }\textbf {\bibinfo {volume} {71}},\ \bibinfo {pages} {022301}
  (\bibinfo {year} {2005})}\BibitemShut {NoStop}%
\bibitem [{\citenamefont {Giampaolo}\ and\ \citenamefont
  {Illuminati}(2010)}]{1367-2630-12-2-025019}%
  \BibitemOpen
  \bibfield  {author} {\bibinfo {author} {\bibfnamefont {S.~M.}\ \bibnamefont
  {Giampaolo}}\ and\ \bibinfo {author} {\bibfnamefont {F.}~\bibnamefont
  {Illuminati}},\ }\href {http://stacks.iop.org/1367-2630/12/i=2/a=025019}
  {\bibfield  {journal} {\bibinfo  {journal} {New Journal of Physics}\ }\textbf
  {\bibinfo {volume} {12}},\ \bibinfo {pages} {025019} (\bibinfo {year}
  {2010})}\BibitemShut {NoStop}%
\bibitem [{\citenamefont {Gualdi}\ \emph {et~al.}(2008)\citenamefont {Gualdi},
  \citenamefont {Kostak}, \citenamefont {Marzoli},\ and\ \citenamefont
  {Tombesi}}]{PhysRevA.78.022325}%
  \BibitemOpen
  \bibfield  {author} {\bibinfo {author} {\bibfnamefont {G.}~\bibnamefont
  {Gualdi}}, \bibinfo {author} {\bibfnamefont {V.}~\bibnamefont {Kostak}},
  \bibinfo {author} {\bibfnamefont {I.}~\bibnamefont {Marzoli}}, \ and\
  \bibinfo {author} {\bibfnamefont {P.}~\bibnamefont {Tombesi}},\ }\href
  {\doibase 10.1103/PhysRevA.78.022325} {\bibfield  {journal} {\bibinfo
  {journal} {Phys. Rev. A}\ }\textbf {\bibinfo {volume} {78}},\ \bibinfo
  {pages} {022325} (\bibinfo {year} {2008})}\BibitemShut {NoStop}%
\bibitem [{\citenamefont {Yao}\ \emph {et~al.}(2011)\citenamefont {Yao},
  \citenamefont {Jiang}, \citenamefont {Gorshkov}, \citenamefont {Gong},
  \citenamefont {Zhai}, \citenamefont {Duan},\ and\ \citenamefont
  {Lukin}}]{PhysRevLett.106.040505}%
  \BibitemOpen
  \bibfield  {author} {\bibinfo {author} {\bibfnamefont {N.~Y.}\ \bibnamefont
  {Yao}}, \bibinfo {author} {\bibfnamefont {L.}~\bibnamefont {Jiang}}, \bibinfo
  {author} {\bibfnamefont {A.~V.}\ \bibnamefont {Gorshkov}}, \bibinfo {author}
  {\bibfnamefont {Z.-X.}\ \bibnamefont {Gong}}, \bibinfo {author}
  {\bibfnamefont {A.}~\bibnamefont {Zhai}}, \bibinfo {author} {\bibfnamefont
  {L.-M.}\ \bibnamefont {Duan}}, \ and\ \bibinfo {author} {\bibfnamefont
  {M.~D.}\ \bibnamefont {Lukin}},\ }\href {\doibase
  10.1103/PhysRevLett.106.040505} {\bibfield  {journal} {\bibinfo  {journal}
  {Phys. Rev. Lett.}\ }\textbf {\bibinfo {volume} {106}},\ \bibinfo {pages}
  {040505} (\bibinfo {year} {2011})}\BibitemShut {NoStop}%
\bibitem [{\citenamefont {Bruderer}\ \emph {et~al.}(2012)\citenamefont
  {Bruderer}, \citenamefont {Franke}, \citenamefont {Ragg}, \citenamefont
  {Belzig},\ and\ \citenamefont {Obreschkow}}]{PhysRevA.85.022312}%
  \BibitemOpen
  \bibfield  {author} {\bibinfo {author} {\bibfnamefont {M.}~\bibnamefont
  {Bruderer}}, \bibinfo {author} {\bibfnamefont {K.}~\bibnamefont {Franke}},
  \bibinfo {author} {\bibfnamefont {S.}~\bibnamefont {Ragg}}, \bibinfo {author}
  {\bibfnamefont {W.}~\bibnamefont {Belzig}}, \ and\ \bibinfo {author}
  {\bibfnamefont {D.}~\bibnamefont {Obreschkow}},\ }\href {\doibase
  10.1103/PhysRevA.85.022312} {\bibfield  {journal} {\bibinfo  {journal} {Phys.
  Rev. A}\ }\textbf {\bibinfo {volume} {85}},\ \bibinfo {pages} {022312}
  (\bibinfo {year} {2012})}\BibitemShut {NoStop}%
\bibitem [{\citenamefont {Paganelli}\ \emph {et~al.}(2013)\citenamefont
  {Paganelli}, \citenamefont {Lorenzo}, \citenamefont {Apollaro}, \citenamefont
  {Plastina},\ and\ \citenamefont {Giorgi}}]{PhysRevA.87.062309}%
  \BibitemOpen
  \bibfield  {author} {\bibinfo {author} {\bibfnamefont {S.}~\bibnamefont
  {Paganelli}}, \bibinfo {author} {\bibfnamefont {S.}~\bibnamefont {Lorenzo}},
  \bibinfo {author} {\bibfnamefont {T.~J.~G.}\ \bibnamefont {Apollaro}},
  \bibinfo {author} {\bibfnamefont {F.}~\bibnamefont {Plastina}}, \ and\
  \bibinfo {author} {\bibfnamefont {G.~L.}\ \bibnamefont {Giorgi}},\ }\href
  {\doibase 10.1103/PhysRevA.87.062309} {\bibfield  {journal} {\bibinfo
  {journal} {Phys. Rev. A}\ }\textbf {\bibinfo {volume} {87}},\ \bibinfo
  {pages} {062309} (\bibinfo {year} {2013})}\BibitemShut {NoStop}%
\bibitem [{\citenamefont {Lorenzo}\ \emph {et~al.}(2013)\citenamefont
  {Lorenzo}, \citenamefont {Apollaro}, \citenamefont {Sindona},\ and\
  \citenamefont {Plastina}}]{PhysRevA.87.042313}%
  \BibitemOpen
  \bibfield  {author} {\bibinfo {author} {\bibfnamefont {S.}~\bibnamefont
  {Lorenzo}}, \bibinfo {author} {\bibfnamefont {T.~J.~G.}\ \bibnamefont
  {Apollaro}}, \bibinfo {author} {\bibfnamefont {A.}~\bibnamefont {Sindona}}, \
  and\ \bibinfo {author} {\bibfnamefont {F.}~\bibnamefont {Plastina}},\ }\href
  {\doibase 10.1103/PhysRevA.87.042313} {\bibfield  {journal} {\bibinfo
  {journal} {Phys. Rev. A}\ }\textbf {\bibinfo {volume} {87}},\ \bibinfo
  {pages} {042313} (\bibinfo {year} {2013})}\BibitemShut {NoStop}%
\bibitem [{\citenamefont {Apollaro}\ \emph {et~al.}(2012)\citenamefont
  {Apollaro}, \citenamefont {Banchi}, \citenamefont {Cuccoli}, \citenamefont
  {Vaia},\ and\ \citenamefont {Verrucchi}}]{PhysRevA.85.052319}%
  \BibitemOpen
  \bibfield  {author} {\bibinfo {author} {\bibfnamefont {T.~J.~G.}\
  \bibnamefont {Apollaro}}, \bibinfo {author} {\bibfnamefont {L.}~\bibnamefont
  {Banchi}}, \bibinfo {author} {\bibfnamefont {A.}~\bibnamefont {Cuccoli}},
  \bibinfo {author} {\bibfnamefont {R.}~\bibnamefont {Vaia}}, \ and\ \bibinfo
  {author} {\bibfnamefont {P.}~\bibnamefont {Verrucchi}},\ }\href {\doibase
  10.1103/PhysRevA.85.052319} {\bibfield  {journal} {\bibinfo  {journal} {Phys.
  Rev. A}\ }\textbf {\bibinfo {volume} {85}},\ \bibinfo {pages} {052319}
  (\bibinfo {year} {2012})}\BibitemShut {NoStop}%
\bibitem [{\citenamefont {Banchi}\ \emph {et~al.}(2011)\citenamefont {Banchi},
  \citenamefont {Apollaro}, \citenamefont {Cuccoli}, \citenamefont {Vaia},\
  and\ \citenamefont {Verrucchi}}]{1367-2630-13-12-123006}%
  \BibitemOpen
  \bibfield  {author} {\bibinfo {author} {\bibfnamefont {L.}~\bibnamefont
  {Banchi}}, \bibinfo {author} {\bibfnamefont {T.~J.~G.}\ \bibnamefont
  {Apollaro}}, \bibinfo {author} {\bibfnamefont {A.}~\bibnamefont {Cuccoli}},
  \bibinfo {author} {\bibfnamefont {R.}~\bibnamefont {Vaia}}, \ and\ \bibinfo
  {author} {\bibfnamefont {P.}~\bibnamefont {Verrucchi}},\ }\href
  {http://stacks.iop.org/1367-2630/13/i=12/a=123006} {\bibfield  {journal}
  {\bibinfo  {journal} {New Journal of Physics}\ }\textbf {\bibinfo {volume}
  {13}},\ \bibinfo {pages} {123006} (\bibinfo {year} {2011})}\BibitemShut
  {NoStop}%
\bibitem [{\citenamefont {Banchi}\ \emph {et~al.}(2010)\citenamefont {Banchi},
  \citenamefont {Apollaro}, \citenamefont {Cuccoli}, \citenamefont {Vaia},\
  and\ \citenamefont {Verrucchi}}]{PhysRevA.82.052321}%
  \BibitemOpen
  \bibfield  {author} {\bibinfo {author} {\bibfnamefont {L.}~\bibnamefont
  {Banchi}}, \bibinfo {author} {\bibfnamefont {T.~J.~G.}\ \bibnamefont
  {Apollaro}}, \bibinfo {author} {\bibfnamefont {A.}~\bibnamefont {Cuccoli}},
  \bibinfo {author} {\bibfnamefont {R.}~\bibnamefont {Vaia}}, \ and\ \bibinfo
  {author} {\bibfnamefont {P.}~\bibnamefont {Verrucchi}},\ }\href {\doibase
  10.1103/PhysRevA.82.052321} {\bibfield  {journal} {\bibinfo  {journal} {Phys.
  Rev. A}\ }\textbf {\bibinfo {volume} {82}},\ \bibinfo {pages} {052321}
  (\bibinfo {year} {2010})}\BibitemShut {NoStop}%
\bibitem [{\citenamefont {Di~Franco}\ \emph {et~al.}(2008)\citenamefont
  {Di~Franco}, \citenamefont {Paternostro},\ and\ \citenamefont
  {Kim}}]{PhysRevLett.101.230502}%
  \BibitemOpen
  \bibfield  {author} {\bibinfo {author} {\bibfnamefont {C.}~\bibnamefont
  {Di~Franco}}, \bibinfo {author} {\bibfnamefont {M.}~\bibnamefont
  {Paternostro}}, \ and\ \bibinfo {author} {\bibfnamefont {M.~S.}\ \bibnamefont
  {Kim}},\ }\href {\doibase 10.1103/PhysRevLett.101.230502} {\bibfield
  {journal} {\bibinfo  {journal} {Phys. Rev. Lett.}\ }\textbf {\bibinfo
  {volume} {101}},\ \bibinfo {pages} {230502} (\bibinfo {year}
  {2008})}\BibitemShut {NoStop}%
\bibitem [{\citenamefont {Markiewicz}\ and\ \citenamefont
  {Wie\ifmmode~\acute{s}\else \'{s}\fi{}niak}(2009)}]{PhysRevA.79.054304}%
  \BibitemOpen
  \bibfield  {author} {\bibinfo {author} {\bibfnamefont {M.}~\bibnamefont
  {Markiewicz}}\ and\ \bibinfo {author} {\bibfnamefont {M.}~\bibnamefont
  {Wie\ifmmode~\acute{s}\else \'{s}\fi{}niak}},\ }\href {\doibase
  10.1103/PhysRevA.79.054304} {\bibfield  {journal} {\bibinfo  {journal} {Phys.
  Rev. A}\ }\textbf {\bibinfo {volume} {79}},\ \bibinfo {pages} {054304}
  (\bibinfo {year} {2009})}\BibitemShut {NoStop}%
\bibitem [{\citenamefont {Godsil}\ \emph {et~al.}(2012)\citenamefont {Godsil},
  \citenamefont {Kirkland}, \citenamefont {Severini},\ and\ \citenamefont
  {Smith}}]{PhysRevLett.109.050502}%
  \BibitemOpen
  \bibfield  {author} {\bibinfo {author} {\bibfnamefont {C.}~\bibnamefont
  {Godsil}}, \bibinfo {author} {\bibfnamefont {S.}~\bibnamefont {Kirkland}},
  \bibinfo {author} {\bibfnamefont {S.}~\bibnamefont {Severini}}, \ and\
  \bibinfo {author} {\bibfnamefont {J.}~\bibnamefont {Smith}},\ }\href
  {\doibase 10.1103/PhysRevLett.109.050502} {\bibfield  {journal} {\bibinfo
  {journal} {Phys. Rev. Lett.}\ }\textbf {\bibinfo {volume} {109}},\ \bibinfo
  {pages} {050502} (\bibinfo {year} {2012})}\BibitemShut {NoStop}%
\bibitem [{\citenamefont {Bayat}\ \emph {et~al.}(2011)\citenamefont {Bayat},
  \citenamefont {Banchi}, \citenamefont {Bose},\ and\ \citenamefont
  {Verrucchi}}]{Bayat2011}%
  \BibitemOpen
  \bibfield  {author} {\bibinfo {author} {\bibfnamefont {A.}~\bibnamefont
  {Bayat}}, \bibinfo {author} {\bibfnamefont {L.}~\bibnamefont {Banchi}},
  \bibinfo {author} {\bibfnamefont {S.}~\bibnamefont {Bose}}, \ and\ \bibinfo
  {author} {\bibfnamefont {P.}~\bibnamefont {Verrucchi}},\ }\href {\doibase
  10.1103/PhysRevA.83.062328} {\bibfield  {journal} {\bibinfo  {journal}
  {Physical Review A}\ }\textbf {\bibinfo {volume} {83}} (\bibinfo {year}
  {2011}),\ 10.1103/PhysRevA.83.062328}\BibitemShut {NoStop}%
\bibitem [{\citenamefont {Liu}\ \emph {et~al.}(2013)\citenamefont {Liu},
  \citenamefont {Guo},\ and\ \citenamefont {Zhou}}]{0295-5075-102-5-50003}%
  \BibitemOpen
  \bibfield  {author} {\bibinfo {author} {\bibfnamefont {Y.}~\bibnamefont
  {Liu}}, \bibinfo {author} {\bibfnamefont {Y.}~\bibnamefont {Guo}}, \ and\
  \bibinfo {author} {\bibfnamefont {D.~L.}\ \bibnamefont {Zhou}},\ }\href
  {http://stacks.iop.org/0295-5075/102/i=5/a=50003} {\bibfield  {journal}
  {\bibinfo  {journal} {EPL (Europhysics Letters)}\ }\textbf {\bibinfo {volume}
  {102}},\ \bibinfo {pages} {50003} (\bibinfo {year} {2013})}\BibitemShut
  {NoStop}%
\bibitem [{\citenamefont {Bowdrey}\ \emph {et~al.}(2002)\citenamefont
  {Bowdrey}, \citenamefont {Oi}, \citenamefont {Short}, \citenamefont
  {Banaszek},\ and\ \citenamefont {Jones}}]{BOS+2002}%
  \BibitemOpen
  \bibfield  {author} {\bibinfo {author} {\bibfnamefont {M.~D.}\ \bibnamefont
  {Bowdrey}}, \bibinfo {author} {\bibfnamefont {D.~K.}\ \bibnamefont {Oi}},
  \bibinfo {author} {\bibfnamefont {A.}~\bibnamefont {Short}}, \bibinfo
  {author} {\bibfnamefont {K.}~\bibnamefont {Banaszek}}, \ and\ \bibinfo
  {author} {\bibfnamefont {J.}~\bibnamefont {Jones}},\ }\href {\doibase
  http://dx.doi.org/10.1016/S0375-9601(02)00069-5} {\bibfield  {journal}
  {\bibinfo  {journal} {Physics Letters A}\ }\textbf {\bibinfo {volume}
  {294}},\ \bibinfo {pages} {258 } (\bibinfo {year} {2002})}\BibitemShut
  {NoStop}%
\bibitem [{\citenamefont {Nielsen}(2002)}]{Nie2002}%
  \BibitemOpen
  \bibfield  {author} {\bibinfo {author} {\bibfnamefont {M.~A.}\ \bibnamefont
  {Nielsen}},\ }\href {\doibase
  http://dx.doi.org/10.1016/S0375-9601(02)01272-0} {\bibfield  {journal}
  {\bibinfo  {journal} {Physics Letters A}\ }\textbf {\bibinfo {volume}
  {303}},\ \bibinfo {pages} {249 } (\bibinfo {year} {2002})}\BibitemShut
  {NoStop}%
\bibitem [{\citenamefont {Di~Franco}\ \emph {et~al.}(2010)\citenamefont
  {Di~Franco}, \citenamefont {Paternostro},\ and\ \citenamefont
  {Kim}}]{PhysRevA.81.022319}%
  \BibitemOpen
  \bibfield  {author} {\bibinfo {author} {\bibfnamefont {C.}~\bibnamefont
  {Di~Franco}}, \bibinfo {author} {\bibfnamefont {M.}~\bibnamefont
  {Paternostro}}, \ and\ \bibinfo {author} {\bibfnamefont {M.~S.}\ \bibnamefont
  {Kim}},\ }\href {\doibase 10.1103/PhysRevA.81.022319} {\bibfield  {journal}
  {\bibinfo  {journal} {Phys. Rev. A}\ }\textbf {\bibinfo {volume} {81}},\
  \bibinfo {pages} {022319} (\bibinfo {year} {2010})}\BibitemShut {NoStop}%
\bibitem [{\citenamefont {Lieb}\ and\ \citenamefont
  {Robinson}(1972)}]{Lieb1972}%
  \BibitemOpen
  \bibfield  {author} {\bibinfo {author} {\bibfnamefont {E.~H.}\ \bibnamefont
  {Lieb}}\ and\ \bibinfo {author} {\bibfnamefont {D.~W.}\ \bibnamefont
  {Robinson}},\ }\href {\doibase 10.1007/BF01645779} {\bibfield  {journal}
  {\bibinfo  {journal} {Communications in Mathematical Physics}\ }\textbf
  {\bibinfo {volume} {28}},\ \bibinfo {pages} {251} (\bibinfo {year}
  {1972})}\BibitemShut {NoStop}%
\bibitem [{\citenamefont {Hastings}(2004)}]{PhysRevLett.93.140402}%
  \BibitemOpen
  \bibfield  {author} {\bibinfo {author} {\bibfnamefont {M.~B.}\ \bibnamefont
  {Hastings}},\ }\href {\doibase 10.1103/PhysRevLett.93.140402} {\bibfield
  {journal} {\bibinfo  {journal} {Phys. Rev. Lett.}\ }\textbf {\bibinfo
  {volume} {93}},\ \bibinfo {pages} {140402} (\bibinfo {year}
  {2004})}\BibitemShut {NoStop}%
\bibitem [{\citenamefont {Guo}\ \emph {et~al.}(2012)\citenamefont {Guo},
  \citenamefont {Liu},\ and\ \citenamefont {Zhou}}]{Guo2012}%
  \BibitemOpen
  \bibfield  {author} {\bibinfo {author} {\bibfnamefont {Y.}~\bibnamefont
  {Guo}}, \bibinfo {author} {\bibfnamefont {Y.}~\bibnamefont {Liu}}, \ and\
  \bibinfo {author} {\bibfnamefont {D.}~\bibnamefont {Zhou}},\ }\href@noop {}
  {\bibfield  {journal} {\bibinfo  {journal} {The European Physical Journal D}\
  }\textbf {\bibinfo {volume} {66}} (\bibinfo {year} {2012})}\BibitemShut
  {NoStop}%
\end{thebibliography}%

\appendix

\section*{Solutions of $F_{N}=0$ in the XY model}

Consider the special case of XY model, where $J_{i}=\frac{1}{2}$,
$K_{i}=\frac{K}{2}$. Then we can rewrite Eq. (\ref{eq:9}) 

\[
\begin{bmatrix}F_{2m}\\
F_{2m-1}
\end{bmatrix}=\begin{bmatrix}p^{2}+K^{2} & p\\
p & 1
\end{bmatrix}\begin{bmatrix}F_{2m-2}\\
F_{2m-3}
\end{bmatrix}
\]
 with 
\[
\begin{bmatrix}F_{0}\\
F_{-1}
\end{bmatrix}=\begin{bmatrix}1\\
0
\end{bmatrix}.
\]

So
\[
\begin{bmatrix}F_{2M}\\
F_{2M-1}
\end{bmatrix}=\mathcal{M}^{M}\begin{bmatrix}1\\
0
\end{bmatrix},
\]
 where 
\[
\mathcal{M}=\begin{bmatrix}p^{2}+K^{2} & p\\
p & 1
\end{bmatrix}.
\]

The eigenvalues of the matrix $\mathcal{M}$ are 
\[
 \frac{1}{2}\left(-\sqrt{\left(-K^{2}-p^{2}-1\right)^{2}-4K^{2}}+K^{2}+p^{2}+1\right)
\]
and 
\[
\frac{1}{2}\left(\sqrt{\left(-K^{2}-p^{2}-1\right)^{2}-4K^{2}}+K^{2}+p^{2}+1\right) .
\]
 Its eigenvectors are 
\[
\begin{bmatrix}-\frac{-K^{2}+\sqrt{K^{4}+2K^{2}p^{2}-2K^{2}+p^{4}+2p^{2}+1}-p^{2}+1}{2p}\\
1
\end{bmatrix}
\]
 and 
\[
\begin{bmatrix}-\frac{-K^{2}-\sqrt{K^{4}+2K^{2}p^{2}-2K^{2}+p^{4}+2p^{2}+1}-p^{2}+1}{2p}\\
1
\end{bmatrix}.
\]
 Note that the eigenvectors are not normalized. After directly compute
and let $p^{2}=-1-2K\cos\varphi-K^{2}$, we get 
\[
\begin{bmatrix}F_{2M}\\
F_{2M-1}
\end{bmatrix}=\begin{bmatrix}
  \scriptstyle (-1)^{\scriptscriptstyle M}K^{\scriptscriptstyle M-1}\csc(\varphi)(K\sin(M+1)\varphi+\sin
  M\varphi)\\
  \scriptstyle (-1)^{M+1}K^{M-1}\csc(\varphi)p\sin(M\varphi)
\end{bmatrix}.
\]

When $N=2M-1$, the roots of the equation $F_{N}=0$ are
$p=\pm{i}\sqrt{1+2K\cos\varphi+K^{2}}$ and $p=0$, where
$\varphi=\frac{k\pi}{M}$, $k=1,\,2,\cdots M-1$.

When $N=2M$, the roots of the equation $F_{N}=0$ are
$p=\pm{i}\sqrt{1+2K\cos\varphi+K^{2}}$, with $\varphi$ is the roots of
the equation
\[
\csc(\varphi)(K\sin(M\varphi+\varphi)-\sin(M\varphi))=0.
\]

For the XX model, that is $K=1$, the roots of the equation $F_{N}=0$
are reduced to
\[
p=\pm i2\cos\frac{\varphi}{2},
\]
 with $\varphi=\frac{2k\pi}{N+1}$, $k=1,\,2,\cdots,\, N$.
 
\section*{Proof of the optimized fidelity}

Here we will prove that for any XY spin chain

\begin{equation*}
  F_{op}(t) = \max_{S,\rho_{2\cdots N}} F(t),
\end{equation*}
where
\begin{equation*}
  F_{op}(t) =  \frac{1}{2} + \frac { \pqty{\abs{a_{N}} +
      \abs{a_{N}^{J_{i}\leftrightarrow K_{i}}}}  
    +  \abs{a_{N} a_{N}^{J_{i}\leftrightarrow K_{i}}} } {6}.
\end{equation*}

Without loss of generality, we prove the result in the case that $N$
is odd. Inserting Eq.~\eqref{eq:5} and Eq.~\eqref{eq:12} into
Eq.~\eqref{eq:3}, we get

\begin{eqnarray*}
  X_{N}(t) & = & a_{N}X_{1}Z_{2}^{N}+\bar{X}_{N},\\
  Y_{N}(t) & = & b_{N}Y_{1}Z_{2}^{N}+\bar{Y}_{N}.
\end{eqnarray*}
Then
\begin{eqnarray*}
  Z_{N}(t) & = & a_{N}b_{N}Z_{1}-ia_{N}X_{1}Z_{2}^{N}\bar{Y}_{N}\\
  &  & -ib_{N}Y_{1}\bar{X}_{N}Z_{2}^{N}-i\bar{X}_{N}\bar{Y}_{N}.\\
\end{eqnarray*}

Therefore
\begin{eqnarray*}
  F(t) & = & \frac{1}{2} + \frac{1}{6} \Big(R_{xx}a_{N}\ev{Z_{2}^{N}}
  + R_{yy}b_{N}\ev{Z_{2}^{N}} + R_{zz}a_{N}b_{N}\\
  &  & + R_{xz}a_{N}\ev{-iZ_{2}^{N}\bar{Y}_{N}} + R_{yz}b_{N}\ev{-i\bar{X}_{N}Z_{2}^{N}}\Big)\\
  & = & \frac{1}{2} + \frac{1}{6} \Big(a_{N} \qty(R_{xx}\ev{Z_{2}^{N}} + R_{xz}\ev{-iZ_{2}^{N}\bar{Y}_{N}})\\
  &  & + b_{N} \qty(R_{yy}\ev{Z_{2}^{N}} +
  R_{yz}\ev{-i\bar{X}_{N}Z_{2}^{N}}) +R_{zz}a_{N}b_{N}\Big) \\
  & \le &
  \frac{1}{2} + \frac{1}{6} \Big(\abs{a_{N}}\sqrt{R_{xx}^{2} + R_{xz}^{2}}\sqrt{\ev{Z_{2}^{N}}^{2}+\ev{-iZ_{2}^{N}
      \bar{Y}_{N}}^{2}}\\
  &  & + \abs{b_{N}} \sqrt{R_{yy}^{2}+R_{yz}^{2}} \sqrt{\ev{Z_{2}^{N}}^{2}+\ev{-iZ_{2}^{N}\bar{X}_{N}}^{2}}\\
  &  & + \abs{a_{N}b_{N}}\abs{R_{zz}}\Big)\\
  & \le & \frac{1}{2} + \frac{1}{6} \qty(\abs{a_{N}}+\abs{b_{N}}+\abs{a_{N}b_{N}}),
\end{eqnarray*}

where we used
\begin{eqnarray*}
  R_{xx}^{2}+R_{xz}^{2} & \le & R_{xx}^{2}+R_{xz}^{2}+R_{xy}^{2}=1,\\
  R_{yy}^{2}+R_{yz}^{2} & \le & R_{yy}^{2}+R_{yz}^{2}+R_{yx}^{2}=1,\\
  R_{zz}^{2} & \le & R_{xz}^{2}+R_{yz}^{2}+R_{zz}^{2}=1.
\end{eqnarray*}

In addition, we define 
\begin{eqnarray*}
  Z_{L} & = & Z_{2}^{N},\\
  X_{L} & = & \frac{-iZ_{2}^{N}\bar{Y}_{N}}{\sqrt{1-b_{N}^{2}}}.
\end{eqnarray*}
It is easy to check that 
\begin{eqnarray*}
  Z_{L} & = & Z_{L}^{\dagger},\\
  X_{L} & = & X_{L}^{\dagger},\\
  Z_{L}^{2} & = & X_{L}^{2}=1,\\
  Z_{L}X_{L} & = & -X_{L}Z_{L}.
\end{eqnarray*}
So $Z_{L}$ and $X_{L}$ can be regarded as the logical qubit Pauli
operators, which implies that
\begin{equation*}
  \sqrt{\ev{Z_{2}^{N}}^{2}+\ev{-iZ_{2}^{N}\bar{Y}_{N}}^{2}} \le \sqrt{\ev{Z_{L}}^{2}+\ev{X_{L}}^{2}}\le1.
\end{equation*}

Similarly, we have
\begin{equation*}
  \sqrt{\ev{Z_{2}^{N}}^{2}+\ev{-iZ_{2}^{N} \bar{X}_{N}}^{2}} \le 1.
\end{equation*}

The maximal fidelity is taken if and only if all the qualities are
satisfied. First, we note that $R_{zz}^{2}=1$. Hence we get
$R_{xz}^{2}=R_{yz}^{2}=0$. So $R_{xx}^{2}=R_{yy}^{2}=1$. Therefore
$\ev{Z_{2}^{N}}^{2}=1$.

\end{document}